\documentclass{SciPost}

\hypersetup{
    colorlinks,
    linkcolor={red!50!black},
    citecolor={blue!50!black},
    urlcolor={blue!80!black}
}

\usepackage[bitstream-charter]{mathdesign}
\DeclareSymbolFont{usualmathcal}{OMS}{cmsy}{m}{n}
\DeclareSymbolFontAlphabet{\mathcal}{usualmathcal}

\fancypagestyle{SPstyle}{
\fancyhf{}
\lhead{\colorbox{scipostblue}{\bf \color{white} ~SciPost Physics Proceedings }}
\rhead{{\bf \color{scipostdeepblue} ~Submission }}

\fancyfoot[C]{\textbf{\thepage}}
}

\usepackage{acronym}
\usepackage{caption}
\usepackage{csquotes}
\usepackage{xspace}
\usepackage{siunitx}
\usepackage{subcaption}
\usepackage{blindtext}

\newcommand{\ttbar}{\ensuremath{t\mkern-2mu\bar{t}}\xspace}
\newcommand{\bbbar}{\ensuremath{b\mkern-2mu\bar{b}}\xspace}
\newcommand{\ccbar}{\ensuremath{c\bar{c}}\xspace}

\newcommand{\slepton}{single-lepton\xspace}
\newcommand{\dilepton}{dilepton\xspace}
\newcommand{\xsec}{cross-section\xspace}
\newcommand{\xsecs}{cross-sections\xspace}

\newcommand{\bjet}{$b$-jet\xspace}
\newcommand{\bjets}{$b$-jets\xspace}
\newcommand*{\bquark}{\ensuremath{b\text{-quark}}\xspace}
\newcommand*{\bquarks}{\ensuremath{b\text{-quarks}}\xspace}
\newcommand{\btag}{\ensuremath{b\text{-tagging}}\xspace}

\newcommand{\cjet}{$c$-jet\xspace}
\newcommand{\cjets}{$c$-jets\xspace}
\newcommand*{\cquark}{\ensuremath{c\text{-quark}}\xspace}
\newcommand*{\cquarks}{\ensuremath{c\text{-quarks}}\xspace}
\newcommand{\ctag}{\ensuremath{c\text{-tagging}}\xspace}
\newcommand{\ctagged}{\ensuremath{c\text{-tagged}}\xspace}
\newcommand{\bctagger}{\texorpdfstring{\ensuremath{b}/\ensuremath{c}}{b/c}-tagger\xspace}

\newcommand*{\ttH}{\ensuremath{\ttbar{}H}\xspace}

\newcommand*{\tttt}{\ensuremath{\ttbar\ttbar}\xspace}
\newcommand*{\Wjets}{\ensuremath{W\text{+\,jets}}\xspace}
\newcommand*{\Zjets}{\ensuremath{Z\text{+\,jets}}\xspace}

\newcommand*{\ttbb}{\ensuremath{\ttbar+\bbbar}\xspace}
\newcommand*{\ttcc}{\ensuremath{\ttbar+\ccbar}\xspace}
\newcommand*{\ttbbin}{\ensuremath{\ttbar+{\geq}2b}\xspace}
\newcommand*{\ttccin}{\ensuremath{\ttbar+{\geq}2c}\xspace}
\newcommand*{\ttb}{\ensuremath{\ttbar+1b}\xspace}
\newcommand*{\ttc}{\ensuremath{\ttbar+1c}\xspace}
\newcommand*{\ttB}{\ensuremath{\ttbar+1B}\xspace}
\newcommand*{\ttC}{\ensuremath{\ttbar+1C}\xspace}

\newcommand*{\ttlight}{\ensuremath{\ttbar+\textrm{light}}\xspace}
\newcommand*{\ttbin}{\ensuremath{\ttbar+{\geq}1b}\xspace}
\newcommand*{\ttcin}{\ensuremath{\ttbar+{\geq}1c}\xspace}

\newcommand*{\sigmattccin}{\ensuremath{\sigma^{\mathrm{fid}}(\ttccin)}\xspace}
\newcommand*{\sigmattc}{\ensuremath{\sigma^{\mathrm{fid}}(\ttc)}\xspace}

\newcommand*{\Rttccininc}{\ensuremath{R^{\mathrm{inc}}_{\ttccin}}\xspace}
\newcommand*{\Rttcinc}{\ensuremath{R^{\mathrm{inc}}_{\ttc}}\xspace}

\newcommand*{\bctightb}{\ensuremath{b@60\%}\xspace}
\newcommand*{\bclooseb}{\ensuremath{b@70\%}\xspace}
\newcommand*{\bctightc}{\ensuremath{c@11\%}\xspace}
\newcommand*{\bcloosec}{\ensuremath{c@22\%}\xspace}

\acrodef{3FS}{three-flavor scheme}
\acrodef{4FS}{four-flavor scheme}
\acrodef{5FS}{five-flavor scheme}
\acrodef{CR}{control region}
\acrodef{EM}{electromagnetic}
\acrodef{FSR}{final-state radiation}
\acrodef{ID}{inner tracking detector}
\acrodef{IP}{interaction point}
\acrodef{ISR}{initial-state radiation}
\acrodef{JES}{jet energy scale}
\acrodef{LHC}{Large Hadron Collider}
\acrodef{LO}{leading order}
\acrodef{MC}{Monte Carlo}
\acrodef{MPI}{multi-parton interaction}
\acrodef{NLO}{next-to-leading order}
\acrodef{NLO+PS}{\ac{NLO} plus parton-shower}
\acrodef{NNLO}{next-to-next-to-leading order}
\acrodef{NNLL}{next-to-next-to-leading-log}
\acrodef{PDF}{parton distribution function}
\acrodef{PS}{parton shower}
\acrodef{QCD}{quantum chromodynamics}
\acrodef{SF}{scale factor}
\acrodef{SM}{Standard Model}
\acrodef{SR}{signal region}
\acrodef{WP}{working point}

\newcommand*{\POWHEGBOX}{\textsc{Powheg\,Box\,2}\xspace}
\newcommand*{\POWHEGBOXRES}{\textsc{Powheg\,Box\,Res}\xspace}
\newcommand*{\PYTHIA}{\textsc{Pythia}\,8\xspace}
\newcommand*{\HERWIG}{\textsc{Herwig}\,7\xspace}
\newcommand*{\MGNLO}{\textsc{MadGraph5\_aMC@NLO}\xspace}
\newcommand*{\OPENLOOPS}{\textsc{OpenLoops}\xspace}

\DeclareSIUnit\barn{b}
\DeclareSIUnit\ifb{\per\femto\barn}
\DeclareSIUnit\pb{\pico\barn}

\usepackage[capitalize, noabbrev]{cleveref}

\begin{document}

\pagestyle{SPstyle}

\begin{center}{\Large \textbf{\color{scipostdeepblue}{%
Measurements of $t\bar{t}$ in association with charm quarks\\at 13 TeV with the ATLAS experiment
}}}\end{center}

\begin{center}\textbf{
Knut Zoch\textsuperscript{1$\star$},
on behalf of the ATLAS Collaboration
}\end{center}

\begin{center}
{\bf 1} Laboratory for Particle Physics and Cosmology, Harvard University,\\Cambridge, Massachusetts 02138, USA
\\[\baselineskip]
$\star$ \href{mailto:kzoch@g.harvard.edu}{\small kzoch@g.harvard.edu}
\end{center}

\definecolor{palegray}{gray}{0.95}
\begin{center}
\colorbox{palegray}{
  \begin{tabular}{rr}
  \begin{minipage}{0.36\textwidth}
    \includegraphics[width=60mm,height=1.5cm]{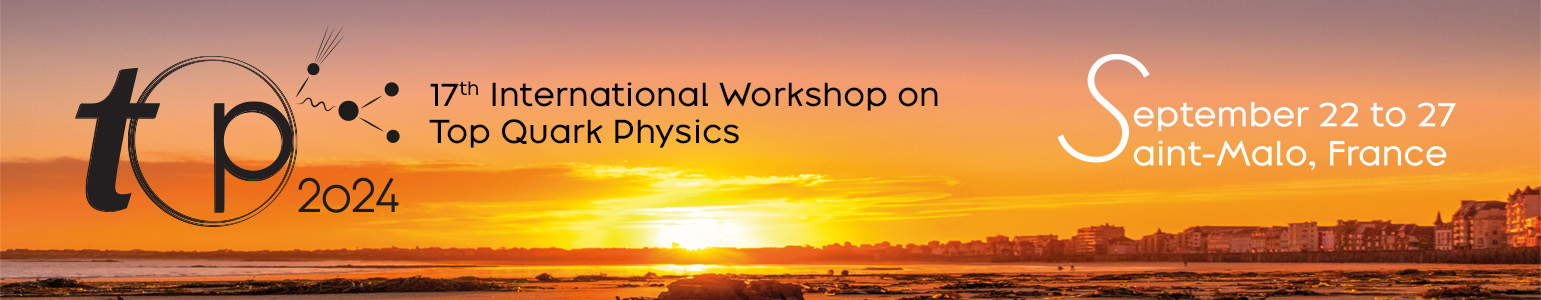}
  \end{minipage}
  &
  \begin{minipage}{0.55\textwidth}
    \begin{center} \hspace{5pt}
    {\it The 17th International Workshop on\\ Top Quark Physics (TOP2024)} \\
    {\it Saint-Malo, France, 22-27 September 2024
    }
    \doi{10.21468/SciPostPhysProc.?}\\
    \end{center}
  \end{minipage}
\end{tabular}
}
\end{center}

\section*{\color{scipostdeepblue}{Abstract}}
\textbf{\boldmath{%
This talk presents the ATLAS Collaboration's first measurement of the inclusive cross-section for top-quark pair production in association with charm quarks.
Using the full Run~2 proton--proton collision data sample at $\sqrt{s} = 13\,\text{TeV}$, collected with the ATLAS experiment at the LHC between 2015 and 2018,
the measurement selects $t\bar{t}$ events with one or two charged leptons and at least one additional jet in the final state.
A custom flavour-tagging algorithm is employed to simultaneously identify $b$-jets and $c$-jets.
The fiducial cross-sections for $t\bar{t}+{\geq}2c$ and $t\bar{t}+1c$ production are found to largely agree with predictions from various $t\bar{t}$ simulations, though all underpredict the observed values.
}}

\vspace{\baselineskip}

\noindent\textcolor{white!90!black}{%
\fbox{\parbox{0.975\linewidth}{%
\textcolor{white!40!black}{\begin{tabular}{lr}%
  \begin{minipage}{0.6\textwidth}%
    {\small Copyright attribution to authors. \newline
    This work is a submission to SciPost Phys. Proc. \newline
    License information to appear upon publication. \newline
    Publication information to appear upon publication.}
  \end{minipage} & \begin{minipage}{0.4\textwidth}
    {\small Received Date \newline Accepted Date \newline Published Date}%
  \end{minipage}
\end{tabular}}
}}
}


\section{Introduction}

\nocite{PERF-2007-01,Evans:2008zzb}

\begin{figure}[tp]
  \centering
  \subcaptionbox{\label{fig:feynman_a}}{%
  \includegraphics[width=0.32\linewidth]{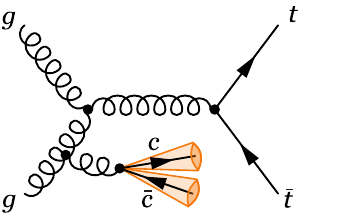}}
  \subcaptionbox{\label{fig:feynman_b}}{%
  \includegraphics[width=0.32\linewidth]{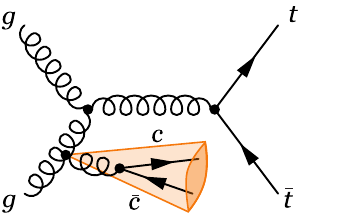}}
  \subcaptionbox{\label{fig:feynman_c}}{%
  \includegraphics[width=0.32\linewidth]{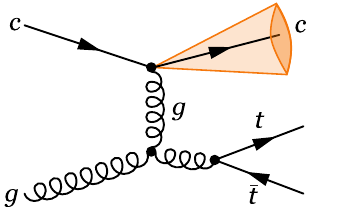}}
  \caption{%
      Illustrative Feynman diagrams for \ttbar pair production with one or multiple additional \cjets:
      \protect\subref{fig:feynman_a} \ttcc production via initial-state gluon radiation where both \cquarks form a jet each;
      \protect\subref{fig:feynman_b} \ttcc production via initial-state gluon radiation where the two \cquarks are in the same jet;
      \protect \subref{fig:feynman_c} \ttc production where the \cquark originates from the initial state.
      Figure taken from Ref.~\cite{TOPQ-2021-26}.
  }
  \label{fig:feynman}
\end{figure}

Top-quark pair production (\ttbar) with additional heavy-flavour jets is a large irreducible background to many other rare processes predicted by the \ac{SM} of particle physics.
Prominent examples include \ttH production with $H \to \bbbar$ decays and the production of four top quarks (\tttt) in single-lepton or dilepton final states.
In this context, \emph{heavy flavour} refers to jets originating from \bquarks or \cquarks.
In \ttbar events, additional heavy-flavour jets can arise from gluon splitting into \bbbar and \ccbar pairs.
The \bbbar or \ccbar pair can form separate jets (\ttbb/\ccbar) or be clustered into a single jet (\ttB/C).
In addition, a single additional \bquark or \cquark can originate from the initial state (\ttb/\ttc).
Illustrative Feynman diagrams for these processes are shown in \cref{fig:feynman}.
Computations of \ttbb production \xsecs exist at \ac{NLO} accuracy in \ac{QCD}, but uncertainties in the choice of the renormalisation and factorisation scales remain sizeable due to the different energy scales involved in the process.
Currently, no dedicated \ttcc computations are available, emphasising the need for experimental measurements to improve the understanding of these final states.

The ATLAS Collaboration~\cite{PERF-2007-01} at the \ac{LHC}~\cite{Evans:2008zzb} at CERN has now performed the first measurement of the inclusive cross-section for \ttbar production in association with charm quarks~\cite{TOPQ-2021-26} to improve the understanding of these final states.
This measurement uses the full Run~2 proton--proton ATLAS dataset at $\sqrt{s} = \SI{13}{\TeV}$ taken between 2015 and 2018, with an integrated luminosity of \SI{140}{\ifb}.
The measurement selects \ttbar events with one or two charged leptons and at least one additional jet in the final state.
A custom flavour-tagging algorithm is employed to simultaneously identify \bjets and \cjets.
For the first time, this measurement then uses regions sensitive to the presence of \cquarks to extract separate fiducial \xsecs for \ttccin and \ttc production.
The first includes events with at least two \cjets, targeting primarily the \ttcc process illustrated in \cref{fig:feynman_a}, while the latter comprises events with one additional \cjet, originating from either the \ttC scenario (\cref{fig:feynman_b}), from \ttc production (\cref{fig:feynman_c}), or from \ttcc where one of the \cjets falls outside the acceptance phase space of the detector.
The CMS Collaboration previously measured \ttcc cross-sections consistent with \ttbar simulation predictions~\cite{CMS-TOP-20-003}.

This document and the associated talk summarise the methodology and results of the new ATLAS measurement~\cite{TOPQ-2021-26}.
Detailed references to all methods, calculations, software tools, etc. are skipped here for brevity but can be found in the original publication.

\section{Simulation of signal and background}

One key component in scrutinising \ttbar production in association with heavy-flavour jets is the modelling of these processes through \ac{MC} simulations.
Recent ATLAS measurements of \ttbb and $\ttH{}(H \to \bbbar)$ production~\cite{TOPQ-2019-03,HIGG-2020-24} have found the modelling of \ttcin to be a limiting factor in the precision of the measurements.
In the absence of dedicated \ttcc computations, inclusive \ttbar{}+jets simulations at \ac{NLO} accuracy interfaced to a \ac{PS} (\acused{NLO+PS}\ac{NLO+PS}) provide the best available model for \ttcin.
Here, the gluon splittings into \ccbar pairs are modelled through the \ac{PS} algorithm, not in the matrix-element calculation.
\ttbin production, on the other hand, can be modelled through \ttbb \ac{4FS} matrix elements, where the \bquark is treated as a massive particle and the production of the \bbbar pair is described directly in the calculation of the hard interaction.

In the ATLAS measurement, \ttbar{}+jets simulations in the \ac{5FS}, generated with \POWHEGBOX~\cite{Frixione:2007nw,Nason:2004rx,Frixione:2007vw,Alioli:2010xd} and the NNPDF3.0\textsc{nlo} \ac{5FS} \ac{PDF} set~\cite{Ball:2014uwa}, are used to model the \ttcin and \ttlight processes.
\ttbb \ac{4FS} simulations using \POWHEGBOXRES~\cite{Jezo:2018yaf}, \OPENLOOPS~\cite{Cascioli:2011va,Denner:2016kdg,Buccioni:2019sur} and the NNPDF3.0\textsc{nlo} \ac{4FS} \ac{PDF} set~\cite{Ball:2014uwa} are employed for the \ttbin process.
Both are interfaced to \PYTHIA~\cite{Sjostrand:2014zea} for the simulation of the \ac{PS} and hadronisation using the ATLAS A14 set of tuned parameters~\cite{ATL-PHYS-PUB-2014-021} and the NNPDF2.3\textsc{lo} \ac{PDF} set~\cite{Ball:2012cx}.
Reweightings and alternative setups (\POWHEGBOX{}+\HERWIG~\cite{Bahr:2008pv,Bellm:2015jjp,Bellm:2017jjp}, varied \ac{NLO} matching) assess modeling uncertainties.
For comparison only, an additional set of \ttbar{}+jets events is generated using \MGNLO~\cite{Alwall:2014hca} interfaced to \HERWIG.

All simulated events are processed through the full ATLAS detector simulation and reconstruction chain, and are then split into four categories: \ttccin, \ttc, \ttbbin, and \ttlight.
The categorisation is based on the presence of additional \bjets and \cjets clustered at the stable particle level before detector simulation.
Those events with one or more \bjets are assigned to the \ttbbin category, those with one \cjet to the \ttc category, and those with at least two \cjets to the \ttccin category.
The latter two classes of events must not contain any \bjets.
All remaining events are assigned to the \ttlight category.
The categorisation is used to combine the \ttbar{}+jets and \ttbb simulations by removing all \ttbin events from the \ac{5FS} \ttbar{}+jets sample and all but the \ttbin events from the \ac{4FS} \ttbb sample, avoiding double-counting.

Various other processes are considered as background to the measurement, including single-top production, associated production of single top quarks or top-quark pairs with a vector boson, and \Wjets, \Zjets and diboson production.
Dedicated \ac{MC} simulations are used to model these processes.
Background contributions from events with fake-lepton signatures are estimated using data-driven methods in the \slepton channel and from \ac{MC} simulations in the \dilepton channel.
Uncertainties in the dominant background processes are assessed through dedicated modelling uncertainties, while conservative normalisation uncertainties are assigned to the minor backgrounds.

\section{Analysis strategy}

\begin{figure}[tp]
  \centering
  \includegraphics[width=0.8\linewidth]{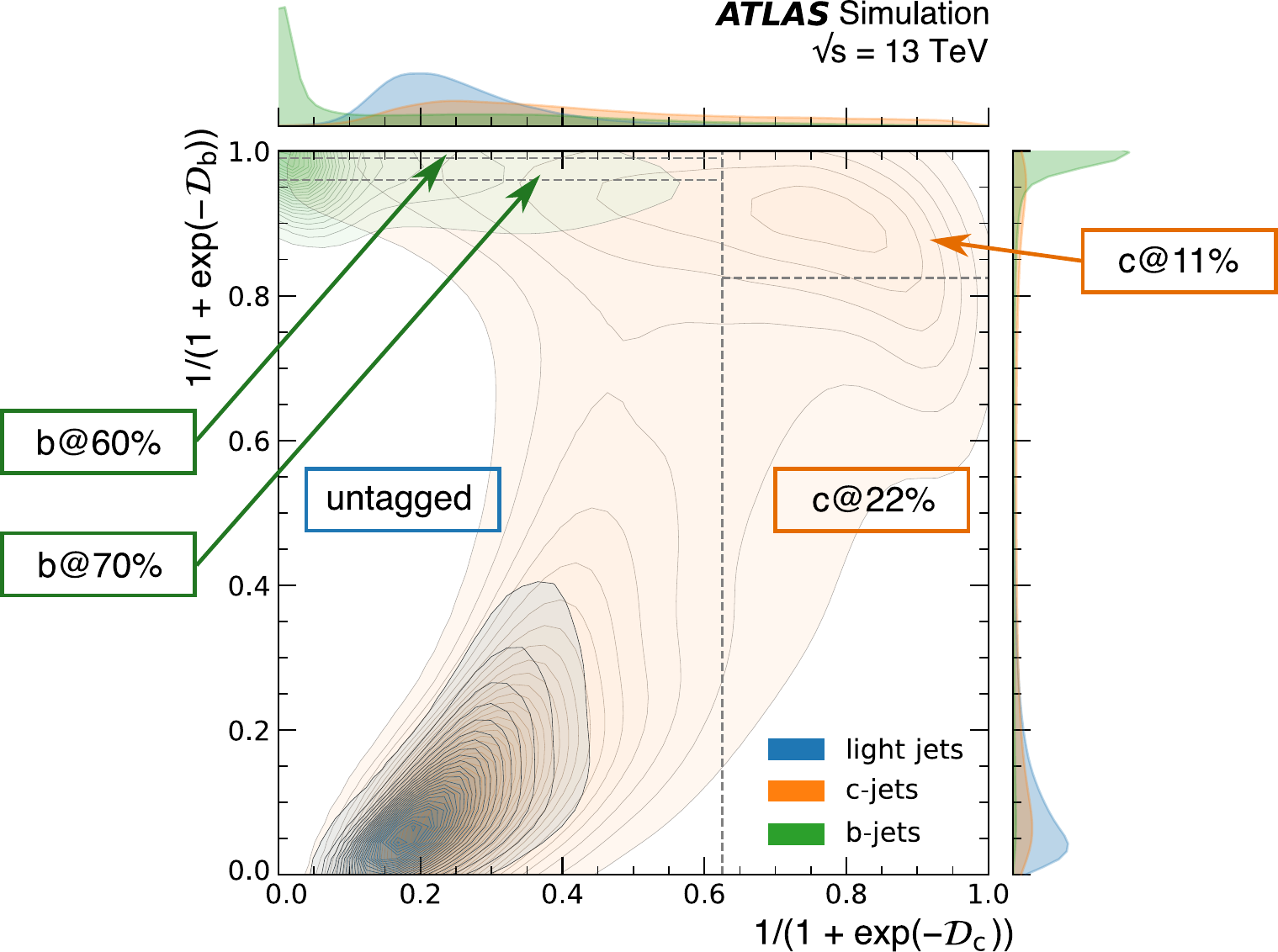}
  \caption{%
      Distribution of light, $c$-, and $b$-jets for the two-dimensional \bctagger in simulated \ttbar events.
      Dashed lines correspond to the edges of the five \bctagger bins.
      For visualisation purposes a standard logistic function is applied to both axes of the discriminant.
      Figure taken from Ref.~\cite{TOPQ-2021-26}.
  }
  \label{fig:bctagger}
\end{figure}

A dedicated flavour-tagging algorithm, the \bctagger, was developed for this analysis to simultaneously identify \bjets and \cjets with high efficiency.
While the standard ATLAS DL1r \btag algorithm~\cite{FTAG-2019-07} provides high efficiency for \bjets, no \ctag working points are available, mandating this custom approach.
The output scores provided by DL1r were re-optimised into a two-dimensional discriminant defining five \bctagger bins: two each for \cjet (\bctightc, \bcloosec) and \bjet (\bctightb, \bclooseb) tagging, and one for untagged jets, as illustrated in \cref{fig:bctagger}.
When using the looser tagging bins, the tighter bins are implicitly included, e.g., the \bcloosec bin includes all jets that pass \bctightc.

The analysis uses a suite of single-electron and single-muon triggers to select events with one or two charged leptons ($e$ or $\mu$), of which one must be matched to the trigger object.
The \slepton channel selects one lepton and $\geq 5$ jets, while the dilepton channel selects two leptons and $\geq 3$ jets, with at least three jets passing \bclooseb or \bcloosec criteria.
Same-flavour lepton pairs are vetoed for low invariant masses and around the $Z$ boson mass.
The \slepton channel is split into events with five jets and events with six or more jets (5-jet-exclusive and 6-jet-inclusive).
The \dilepton channel is split into 3-jet-exclusive and 4-jet-inclusive events.

The five \bctagger bins define 19 orthogonal \acp{CR} and \acp{SR} across the \slepton and \dilepton channels, providing sensitivity to \ttccin, \ttc, and controlling background contributions.
The twelve \acp{CR} are designed to either exclude \ctagged jets or select events with exactly one \ctagged jet, isolating \ttbin, \ttlight, and other background contributions.
The seven \acp{SR} require at least two (one) jet tagged with \bcloosec in the \slepton (\dilepton) channel and have different predicted purities of \ttccin and \ttc events.
These regions are then used to extract the \ttccin and \ttc signal strengths from the data using a profile likelihood fit.
Additionally, normalisation factors for the \ttbin and \ttlight background contributions are included in the fit and are left free-floating.
Systematic uncertainties relating to the modelling of the signal and background processes, the dedicated flavour-tagging algorithm, and all other instrumental uncertainties are included in the fit as nuisance parameters constrained by Gaussian penalty terms.
While the \acp{CR} are included as a single bin in the fit, the jet-inclusive and jet-exclusive \acp{SR} use jet-multiplicity and invariant-mass distributions of the \ctagged jets, respectively.

The measurements are performed in a fiducial phase defined at the stable particle level that mimics the selection criteria of the analysis.
Apart from the lepton requirements, the fiducial phase requires at least one additional jet, but no explicit requirements are imposed on its flavour.
The measurements are also performed in a more inclusive volume without requirements on the \ttbar decay products and the jet multiplicity.

\section{Results}

Good post-fit agreement between data and simulation is observed in all \acp{CR} and \acp{SR} of the analysis.
The goodness of fit was evaluated using a \emph{saturated model} and the compatibility with data was found to be \SI{98}{\percent}.
Using the extracted \ttccin and \ttc signal strengths, the fiducial \xsecs are determined to be
\begin{align}
  \sigmattccin
  & = \num{1.28} \, ^{+0.16}_{-0.10} \, \text{(stat)} \, ^{+0.21}_{-0.22} \, \text{(syst)} \; \unit{\pb}
  = 1.28 \, ^{+0.27}_{-0.24} \; \unit{\pb},\\
  \sigmattc
  &= \num{6.4} \, ^{+0.5}_{-0.4} \, \text{(stat)} \, \pm 0.8 \, \text{(syst)} \; \unit{\pb}
  = 6.4 \, ^{+1.0}_{-0.9} \; \unit{\pb}.
\end{align}
The largest contributing sources of uncertainties are the modelling of \ttcin, \ttbin, and \ttlight, in particular in the \ac{NLO} matching and the \ac{PS}, the uncertainties in the \bctagger, and the data statistics.
The measured \ttbin normalisation factor is compatible with those obtained in the dedicated ATLAS \ttbb measurements~\cite{TOPQ-2019-03}.

\begin{figure}[t]
  \centering
  \includegraphics[width=0.98\linewidth]{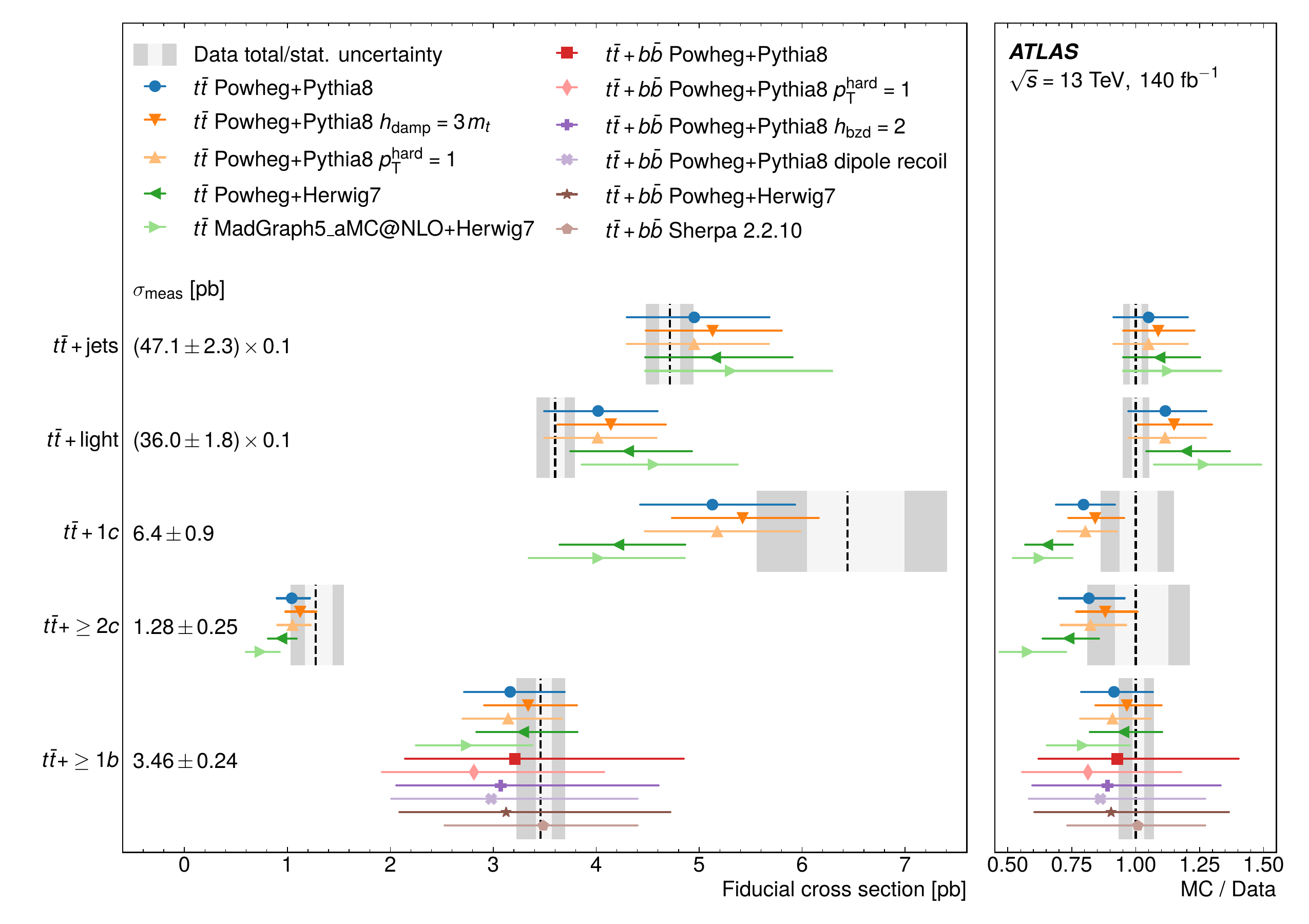}
  \caption{%
      Measured fiducial cross-section values in comparison with various \ac{NLO+PS} predictions from inclusive \ttbar{}+jets and \ttbb simulations.
      Figure taken from Ref.~\cite{TOPQ-2021-26}.
  }
  \label{fig:enter-label}
\end{figure}

The measured values are compared to various \ac{NLO+PS} predictions from inclusive \ttbar{}+jets and \ttbb simulations in \cref{fig:enter-label}.
The prediction for \ttccin and \ttc are largely consistent with the measured values, though all predictions underpredict the observed values.
Considering scale and \ac{PDF} variations on the predictions, the \ttbar \POWHEGBOX{}+\PYTHIA{} predictions agree within 1.1 standard deviations of measurement and prediction uncertainties.
Agreement between 1.2 and 2.0 standard deviations is observed for the \POWHEGBOX{}+\HERWIG and the \MGNLO{}+\HERWIG predictions.
Fit stability tests, including combined and independent \ttccin and \ttc parameterisations, confirm robustness.

In addition to the fiducial \xsecs, the ratios of \ttccin and \ttc to total \ttbar{}+jets production are extracted in the fiducial and the more inclusive volume.
The values were found to be $\Rttccininc = (1.23 \pm 0.25) \%$ and $\Rttcinc = (8.8 \pm 1.3) \%$ in inclusive volume, in agreement with \ac{NLO+PS} simulations at a similar level as the fiducial \xsecs.

\section{Conclusion}

The ATLAS Collaboration at the \ac{LHC} has performed their first measurement of the inclusive cross-section for \ttbar production in association with charm quarks~\cite{TOPQ-2021-26}.
The measurement uses the full Run~2 proton--proton collision data sample at $\sqrt{s} = \SI{13}{\TeV}$, corresponding to an integrated luminosity of \SI{140}{\ifb}, collected between 2015 and 2018.
Using \slepton and \dilepton events and a custom flavour-tagging algorithm to identify \bjets and \cjets, the fiducial \xsecs for \ttccin and \ttc production are extracted separately from the data for the first time.
The measured values are found to be in agreement with various \ac{NLO+PS} predictions from inclusive \ttbar{}+jets and \ttbb simulations, though all predictions underpredict the observed values.
Precision is limited by modelling uncertainties (\ttcin, \ttbin, \ttlight), uncertainties in the \bctagger, and data statistics.
Measurements of the ratios of the \ttccin and \ttc \xsecs to the inclusive \ttbar{}+jets \xsec are also performed and found to be in agreement with the predictions from \ac{NLO+PS} simulations at a similar level as the fiducial \xsecs.

\section*{Acknowledgements}


The author received support from the U.S. Department of Energy Award No. DE-SC0007881.

\vspace{1\baselineskip}\noindent
Copyright 2024 CERN for the benefit of the ATLAS Collaboration. Reproduction of this article or parts of it is allowed as specified in the CC-BY-4.0 license.

\bibliography{bibfile.bib}

\begin{thebibliography}{10}
\providecommand{\url}[1]{\texttt{#1}}
\providecommand{\urlprefix}{URL }
\expandafter\ifx\csname urlstyle\endcsname\relax
  \providecommand{\doi}[1]{doi:\discretionary{}{}{}#1}\else
  \providecommand{\doi}{doi:\discretionary{}{}{}\begingroup
  \urlstyle{rm}\Url}\fi
\providecommand{\eprint}[2][]{\url{#2}}

\bibitem{PERF-2007-01}
{ATLAS Collaboration},
\newblock \emph{{The ATLAS Experiment at the CERN Large Hadron Collider}},
\newblock JINST \textbf{3}, S08003 (2008),
\newblock \doi{10.1088/1748-0221/3/08/S08003}.

\bibitem{Evans:2008zzb}
L.~Evans and P.~Bryant,
\newblock \emph{{LHC Machine}},
\newblock JINST \textbf{3}, S08001 (2008),
\newblock \doi{10.1088/1748-0221/3/08/S08001}.

\bibitem{TOPQ-2021-26}
{ATLAS Collaboration},
\newblock \emph{{Measurement of top-quark pair production in association with
  charm quarks in proton-proton collisions at $\sqrt{s}=13$ TeV with the ATLAS
  detector}} (2024), \eprint{2409.11305}.

\bibitem{CMS-TOP-20-003}
{CMS Collaboration},
\newblock \emph{{First measurement of the cross section for top quark pair
  production with additional charm jets using dileptonic final states in \(pp\)
  collisions at \(\sqrt{s} = 13\,\text{TeV}\)}},
\newblock Phys. Lett. B \textbf{820}, 136565 (2021),
\newblock \doi{10.1016/j.physletb.2021.136565},
\newblock \eprint{2012.09225}.

\bibitem{TOPQ-2019-03}
{ATLAS Collaboration},
\newblock \emph{{Measurement of $t\bar{t}$ production in association with
  additional $b$-jets in the $e\mu$ final state in proton-proton collisions at
  $\sqrt{s}$=13 TeV with the ATLAS detector}} (2024), \eprint{2407.13473}.

\bibitem{HIGG-2020-24}
{ATLAS Collaboration},
\newblock \emph{{Measurement of the associated production of a
  top-antitop-quark pair and a Higgs boson decaying into a $b\bar{b}$ pair in
  $pp$ collisions at $\sqrt{s}=13$ TeV using the ATLAS detector at the LHC}}
  (2024), \eprint{2407.10904}.

\bibitem{Frixione:2007nw}
S.~Frixione, G.~Ridolfi and P.~Nason,
\newblock \emph{{A positive-weight next-to-leading-order Monte Carlo for heavy
  flavour hadroproduction}},
\newblock JHEP \textbf{09}, 126 (2007),
\newblock \doi{10.1088/1126-6708/2007/09/126},
\newblock \eprint{0707.3088}.

\bibitem{Nason:2004rx}
P.~Nason,
\newblock \emph{{A new method for combining NLO QCD with shower Monte Carlo
  algorithms}},
\newblock JHEP \textbf{11}, 040 (2004),
\newblock \doi{10.1088/1126-6708/2004/11/040},
\newblock \eprint{hep-ph/0409146}.

\bibitem{Frixione:2007vw}
S.~Frixione, P.~Nason and C.~Oleari,
\newblock \emph{{Matching NLO QCD computations with parton shower simulations:
  the POWHEG method}},
\newblock JHEP \textbf{11}, 070 (2007),
\newblock \doi{10.1088/1126-6708/2007/11/070},
\newblock \eprint{0709.2092}.

\bibitem{Alioli:2010xd}
S.~Alioli, P.~Nason, C.~Oleari and E.~Re,
\newblock \emph{{A general framework for implementing NLO calculations in
  shower Monte Carlo programs: the POWHEG BOX}},
\newblock JHEP \textbf{06}, 043 (2010),
\newblock \doi{10.1007/JHEP06(2010)043},
\newblock \eprint{1002.2581}.

\bibitem{Ball:2014uwa}
{NNPDF Collaboration}, R.~D. Ball \emph{et~al.},
\newblock \emph{{Parton distributions for the LHC run II}},
\newblock JHEP \textbf{04}, 040 (2015),
\newblock \doi{10.1007/JHEP04(2015)040},
\newblock \eprint{1410.8849}.

\bibitem{Jezo:2018yaf}
T.~Ježo, J.~M. Lindert, N.~Moretti and S.~Pozzorini,
\newblock \emph{{New NLOPS predictions for \(t\bar{t}+b\)-jet production at the
  LHC}},
\newblock Eur. Phys. J. C \textbf{78}(6), 502 (2018),
\newblock \doi{10.1140/epjc/s10052-018-5956-0},
\newblock \eprint{1802.00426}.

\bibitem{Cascioli:2011va}
F.~Cascioli, P.~Maierh{\"o}fer and S.~Pozzorini,
\newblock \emph{{Scattering Amplitudes with Open Loops}},
\newblock Phys. Rev. Lett. \textbf{108}, 111601 (2012),
\newblock \doi{10.1103/PhysRevLett.108.111601},
\newblock \eprint{1111.5206}.

\bibitem{Denner:2016kdg}
A.~Denner, S.~Dittmaier and L.~Hofer,
\newblock \emph{{\textsc{Collier}: A fortran-based complex one-loop library in
  extended regularizations}},
\newblock Comput. Phys. Commun. \textbf{212}, 220 (2017),
\newblock \doi{10.1016/j.cpc.2016.10.013},
\newblock \eprint{1604.06792}.

\bibitem{Buccioni:2019sur}
F.~Buccioni, J.-N. Lang, J.~M. Lindert, P.~Maierh{\"o}fer, S.~Pozzorini,
  H.~Zhang and M.~F. Zoller,
\newblock \emph{{OpenLoops 2}},
\newblock Eur. Phys. J. C \textbf{79}(10), 866 (2019),
\newblock \doi{10.1140/epjc/s10052-019-7306-2},
\newblock \eprint{1907.13071}.

\bibitem{Sjostrand:2014zea}
T.~Sj{\"o}strand, S.~Ask, J.~R. Christiansen, R.~Corke, N.~Desai, P.~Ilten,
  S.~Mrenna, S.~Prestel, C.~O. Rasmussen and P.~Z. Skands,
\newblock \emph{{An introduction to PYTHIA 8.2}},
\newblock Comput. Phys. Commun. \textbf{191}, 159 (2015),
\newblock \doi{10.1016/j.cpc.2015.01.024},
\newblock \eprint{1410.3012}.

\bibitem{ATL-PHYS-PUB-2014-021}
{ATLAS Collaboration},
\newblock \emph{{ATLAS Pythia~8 tunes to \(7~\text{TeV}\) data}},
\newblock {ATL-PHYS-PUB-2014-021} (2014).

\bibitem{Ball:2012cx}
{NNPDF Collaboration}, R.~D. Ball \emph{et~al.},
\newblock \emph{{Parton distributions with LHC data}},
\newblock Nucl. Phys. B \textbf{867}, 244 (2013),
\newblock \doi{10.1016/j.nuclphysb.2012.10.003},
\newblock \eprint{1207.1303}.

\bibitem{Bahr:2008pv}
M.~B{\"a}hr \emph{et~al.},
\newblock \emph{{Herwig++ physics and manual}},
\newblock Eur. Phys. J. C \textbf{58}, 639 (2008),
\newblock \doi{10.1140/epjc/s10052-008-0798-9},
\newblock \eprint{0803.0883}.

\bibitem{Bellm:2015jjp}
J.~Bellm \emph{et~al.},
\newblock \emph{{Herwig 7.0/Herwig++ 3.0 release note}},
\newblock Eur. Phys. J. C \textbf{76}(4), 196 (2016),
\newblock \doi{10.1140/epjc/s10052-016-4018-8},
\newblock \eprint{1512.01178}.

\bibitem{Bellm:2017jjp}
J.~Bellm \emph{et~al.},
\newblock \emph{{Herwig 7.1 Release Note}}  (2017),
\newblock \eprint{1705.06919}.

\bibitem{Alwall:2014hca}
J.~Alwall, R.~Frederix, S.~Frixione, V.~Hirschi, F.~Maltoni, O.~Mattelaer,
  H.~S. Shao, T.~Stelzer, P.~Torrielli and M.~Zaro,
\newblock \emph{{The automated computation of tree-level and next-to-leading
  order differential cross sections, and their matching to parton shower
  simulations}},
\newblock JHEP \textbf{07}, 079 (2014),
\newblock \doi{10.1007/JHEP07(2014)079},
\newblock \eprint{1405.0301}.

\bibitem{FTAG-2019-07}
{ATLAS Collaboration},
\newblock \emph{{ATLAS flavour-tagging algorithms for the LHC Run~2 \(pp\)
  collision dataset}},
\newblock Eur. Phys. J. C \textbf{83}, 681 (2023),
\newblock \doi{10.1140/epjc/s10052-023-11699-1},
\newblock \eprint{2211.16345}.

\end{thebibliography}

\end{document}